\documentclass[english,12pt]{article}%,draft]{article}
\usepackage[latin9]{inputenc}
\usepackage{amsmath,amssymb}
\usepackage{graphicx}
\usepackage{babel}

\newcommand{\be}{\begin{equation}}
\newcommand{\ee}{\end{equation}}
\newcommand{\bea}{\begin{eqnarray}}
\newcommand{\eea}{\end{eqnarray}}

\def\rt{\tilde{r}}
\def\rhot{\tilde{\rho}}
\newcount\figno
\def\fig#1#2#3{
\par\begingroup\parindent=0pt\leftskip=1cm\rightskip=1cm\parindent=0pt
\baselineskip=11pt
\global\advance\figno by 1
\midinsert
\epsfxsize=#3
\centerline{\epsfbox{#2}}
\vskip 12pt
{\bf Fig.\ \the\figno: } #1\par
\endinsert\endgroup\par
}
\def\figlabel#1{\xdef#1{\the\figno}}
\def\encadremath#1{\vbox{\hrule\hbox{\vrule\kern8pt\vbox{\kern8pt
\hbox{$\displaystyle #1$}\kern8pt}
\kern8pt\vrule}\hrule}}
\def\p{\partial}

\def\D7b{\overline{D7}}

\def\Ac{{\cal A}}
\def\Bc{{\cal B}}
\def\Cc{{\cal C}}
\def\Bh{\hat{B}}
\def\rhoh{\hat{\rho}}
\def\Cch{\hat{\cal C}}
\def\ah{\hat{a}}
\def\omh{\hat{\omega}}

\begin{document}

%\noLabels
%\nobbibitem
%title
%\rightline{arXiv:YYMM.NNNN}

\rightline{HIP-2009-15/TH}
\vskip 2cm \centerline{\large {\bf
AC Transport at Holographic Quantum Hall Transitions }}
%\vskip 0.3cm
%\centerline{\large {\bf in a dual model of quantum phase transition}}
\vskip 1cm

\renewcommand{\thefootnote}{\fnsymbol{footnote}}

\centerline{{\bf Janne Alanen,$^{1,2}$\footnote{janne.alanen@helsinki.fi} Esko Keski-Vakkuri,$^{1}$\footnote{esko.keski-vakkuri@helsinki.fi}
Per Kraus,$^{3}$\footnote{pkraus@ucla.edu}
and Ville Suur-Uski$^{1,2}$\footnote{ville.suur-uski@helsinki.fi} }}
\vskip .5cm \centerline{\it ${}^{1}$Helsinki Institute of Physics
and ${}^{2}$Department of Physics } \centerline{\it
P.O.Box 64, FIN-00014 University of Helsinki, Finland}
\centerline{\it ${}^{3}$ Department of Physics and Astronomy, UCLA,
Los Angeles, CA 90095-1547, USA}

\setcounter{footnote}{0}

\renewcommand{\thefootnote}{\arabic{footnote}}

\begin{abstract}

We compute AC electrical transport at quantum Hall critical points, as modeled
by intersecting branes and gauge/gravity duality.
We compare our results with a previous field theory computation by Sachdev, and find
unexpectedly good agreement. We also give general results for DC Hall and longitudinal conductivities valid for a wide
class of quantum Hall transitions, as well as (semi)analytical results for AC
quantities in special limits.  Our results exhibit a surprising degree of universality;
for example, we find that the high frequency behavior, including subleading behavior, is
identical for our entire class of theories.

\end{abstract}

\newpage

\section{Introduction and Summary}

Not too long ago, it would have seemed highly unlikely that results from quantum gravity and string theory would find their way into discussions of condensed matter physics.  This began to change with an improved understanding of non-perturbative string theory, and especially in the past few years with the steadily growing appreciation of the usefulness of AdS/CFT duality  for studying strongly coupled quantum field theories, examples of which abound in
condensed matter physics.  It is natural to try to forge this connection in systems exhibiting
some degree of universality, which has led to a focus on the physics near quantum critical points.  The quantum critical point itself is achieved by going to zero temperature and tuning the appropriate coupling constants; however its existence controls the physics in a finite
neighborhood transition region that spreads from it.  Similarly, the dual gravity theory can have an AdS  description that extends to the transition region.
The dual description can then be used to  study transport at non-zero temperature near the critical point, among other things.  This is helpful when it is difficult to perform standard calculations based on traditional models of interacting quasiparticles.
For more discussion, see {\em e.g.} the recent reviews \cite{Hartnoll:2009sz,Herzog:2009xv,Sachdev:2008ba}.

In this paper we consider  AdS dual descriptions of  quantum phase transitions corresponding
to transitions between different
(integer or fractional) quantum Hall plateaus, modeled by a system of
intersecting branes \cite{Davis:2008nv} (see also \cite{rey}). These branes typically have a single mutually transverse
direction,   and the transition
is realized by taking the brane separation to zero,  and then continuing through to the other side.   We will give analytical results for DC electrical conductivities valid for a wide class of transitions.
In the middle of the transition ({\it i.e.} at the critical point),
we will compute the longitudinal and transverse AC conductivities at finite temperature. We will then compare the results to those
obtained by Sachdev \cite{sachdev}, who studied finite temperature transport near a fractional
quantum Hall critical point in a simplified field theoretic toy model proposed by Chen, Fisher and Wu
\cite{Chen:1993cd}.  The model has Dirac fermions
coupled to a U(1) Chern-Simons gauge field, converting the fermions to anyons representing
the quasiparticles  of a fractional quantum Hall state. The system undergoes a quantum transition to an insulator when the mass of  the fermions is taken to zero and its sign is reversed, much like what happens in the intersecting brane model.\footnote{In the latter, the transition to zero conductivity is also easily obtained using multiple branes.}

In general, a system undergoing a quantum phase transition has two
qualitatively different regimes of charge transport \cite{damlesachdev, KB}, depending on the ratio of the frequency
of the current and the temperature, $\omh \equiv \hbar\omega/k_BT$:
i) collision-dominated low frequency regime $\omh \ll 1$, and ii) phase-coherent high frequency regime $\omh\gg 1$. In the latter,  charge carriers are excited by the external
perturbation, and do not collide with thermal excitations.
A characteristic feature is that at the scale invariant quantum fixed point  the dynamic conductivity becomes a universal
function of $\omh$ \cite{damlesachdev}. In $2+1$ dimensions (which we focus on),
\be
 \sigma (\omega) = \sigma_0 %\left(\frac{k_BT}{\hbarc}\right)^{(d-2)/z}
 \Sigma (\omh) \ ,
\ee
where $\sigma_0$ is the quantum unit of conductance ($\sigma_0 = e_*^2/h$ where $e_*$ is the carrier charge)
%$d$ is the space dimension, $z$ is the dynamic exponent, and
and $\Sigma(\omh)$ is the universal function. In the limit $\omh\rightarrow \infty$, corresponding
either to the high frequency limit $\omega\rightarrow \infty$ at finite $T$, or the zero temperature limit
$T\rightarrow 0$ at finite frequency, the conductivity approaches the universal conductivity
$\sigma_0\Sigma (\infty)$, where $\Sigma (\infty)$ is a pure number, independent of the microscopic
details of the system, and can be used to classify quantum critical points.   In our dual
models this limiting value will be related to an effective brane tension.

\begin{figure}
\begin{picture}(250,190)
\put(110,190){\includegraphics[scale=0.3, angle=270]{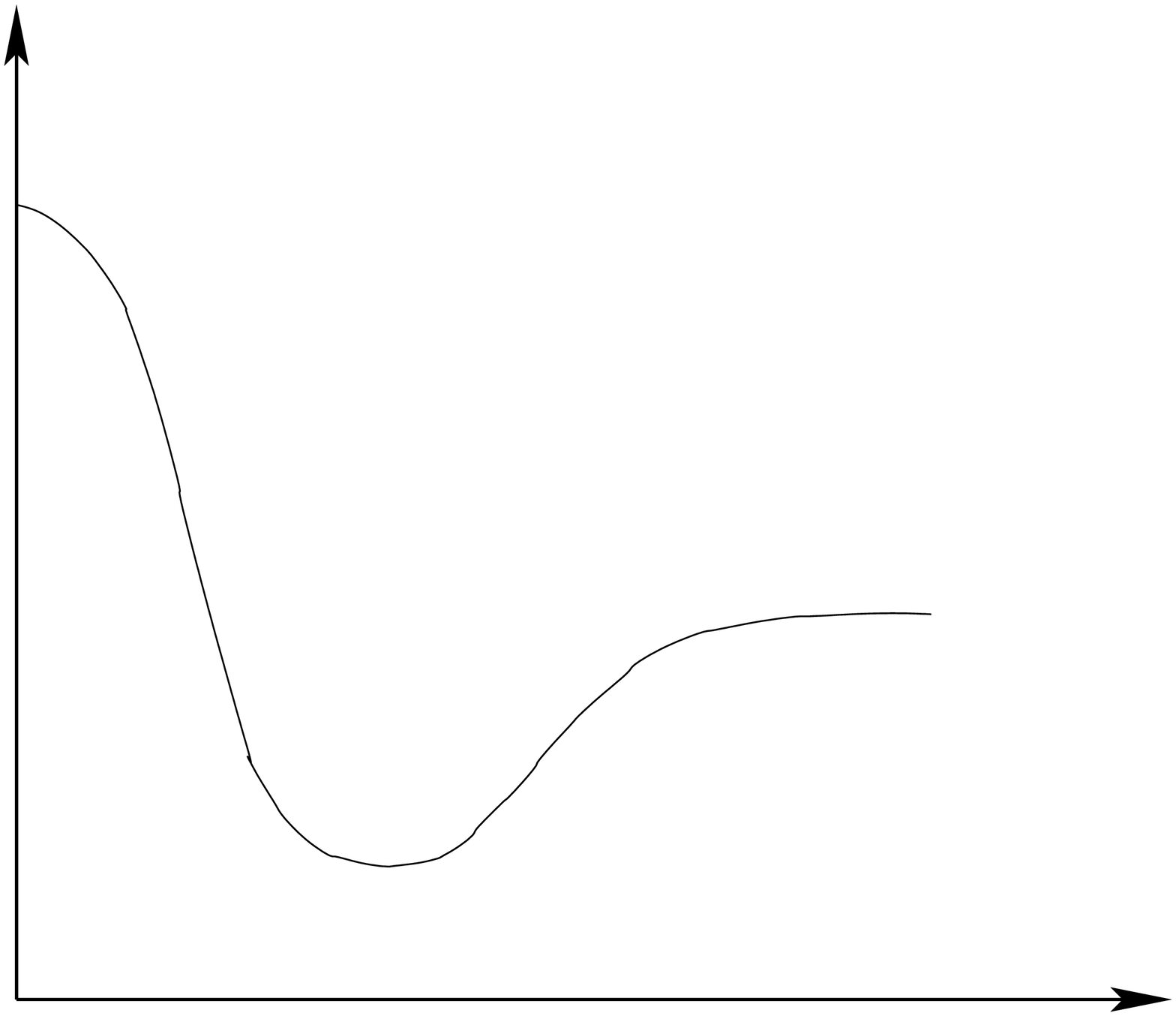}}
\put(290,0){$\omh$}
\put(110,180){$\Sigma'$}
\end{picture}
%kuvateksti
\caption{$\Sigma'$ as a function of $\omh$}
\label{fig1}
\end{figure}

The expected characteristic behavior of the real part of the
complex valued function $\Sigma$, often denoted by $\Sigma'$, is sketched in Figure \ref{fig1} (following \cite{damlesachdev}). It depicts a Drude peak at small $\omh$, and in
the collisionless regime $\omh\gg 1$ the asymptotic behavior of $\Sigma'$ is determined by the scale invariance at the critical point. In particular, the conductivity is expected to become
temperature independent in the high frequency limit, and then dimensional analysis forces
$\Sigma'(\omh)$ to go to a constant in this limit.  In our model, the conductivity at
the finite temperature critical point is actually frequency independent; this is the same
phenomenon as was observed in \cite{Herzog:2007ij}, and for the same underlying reason.
In order to get interesting frequency dependence we further deform the theory by turning
on a nonzero charge density and magnetic field.
Upon so doing,
we will find that  the characteristic behavior of the conductivity agrees with the expectation
of Fig. \ref{fig1}.

\begin{figure}
%maaritetaan kuvan koko (leveys,korkeus)
\centering
\includegraphics[scale=0.8]{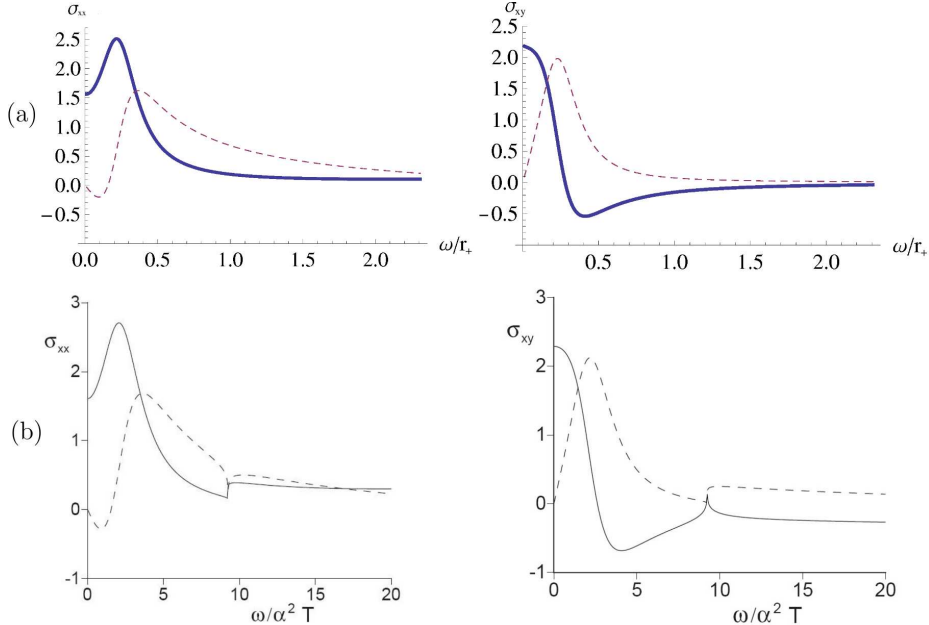}
\caption{Longitudinal conductivity $\sigma_{xx}$
 and Hall conductivity $\sigma_{xy}$ a) from the D-brane model and  b) from Sachdev's analysis \cite{sachdev}. Real parts are depicted by the solid curves and imaginary parts by the dashed lines.
 The plots in a) use parameter values $\rhoh=15.1,\Bh=1.4,\tau=0.3$ of the model.}
 \label{fig2}
\end{figure}

In particular, we will compare our results against those of Sachdev \cite{sachdev}, and find
unexpectedly good agreement.
For the comparison, it is interesting to note that in \cite{sachdev} the results come from various
detailed calculations. The Hall conductivity is given entirely by perturbative contributions, reducing
to coherent transport of externally created particle-hole pairs, while the longitudinal component also receives a contribution from incoherent collision effects \cite{sachdev}. The analysis is quite
tricky even in the simplified model, and some discontinuities remain as artifacts in the final plots
of the conductivity. Given the different ingredients that go into the analysis, it is quite remarkable that we find the dual D-brane construction to give such a good match (fitting just two parameters already
gives a good fit to the overall shape, and the third parameter fits the asymptotic behavior) to the results of \cite{sachdev}.  The comparison  is presented in Figures \ref{fig2} (a) and (b).  On the other hand, it should be kept in mind that there are some
important differences. Namely, the model studied in  \cite{sachdev} is a scale invariant theory,
while we are breaking scale invariance by including a nonzero charge density and magnetic field.

As in other AdS setups, numerical methods are  needed to compute the AC conductivities for
general values of parameters.  However, we are also able to to find some analytical results
in special cases.  One case is for the DC conductivity, where we can compute the Hall and
longitudinal conductivities during the transition for general values of charge and magnetic
fields.   Another example is that we can give an almost analytical expression ({\it i.e.} in terms of an integral) for the AC response at the transition point in the presence of a small charge and magnetic field.  We are not aware of other cases where it is possible to find such a result.  We can also give an analytical derivation of the response in the
high frequency limit, including the  subleading behavior in $1/\omega$.   All of these
analytical results of course agree with our numerics in the appropriate regime.

Our calculations follow the logic of earlier work on dual models of charge transport, see {\it e.g.} \cite{Herzog:2007ij,Hartnoll:2007ai,Hartnoll:2007ih,Hartnoll:2007ip,Myers:2008me} and the reviews cited above.
%\cite{Herzog:2007ij,Myers:2007we,Hartnoll:2007ai,Hartnoll:2007ih,Karch:2007pd,Hartnoll:2007ip,O'Bannon:2007in,
%Myers:2008zz,Hartnoll:2008vx,Gubser:2008wz,Gubser:2008pf,Horowitz:2008bn}.
Studies of the quantum Hall effect in string theory include the early work \cite{Brodie:2000yz,Bergman:2001qg,Hellerman:2001yv},
a recent AdS dual model \cite{KeskiVakkuri:2008eb} which utilizes the construction in \cite{Hartnoll:2008vx}, and the most recent constructions \cite{Fujita:2009kw,Hikida:2009tp} inspired by the ABJM \cite{Aharony:2008ug} model for M2-brane dynamics.

\section{Quantum Hall critical points from intersecting branes}

A quantum Hall plateau transition describes a jump in the transverse conductivity
$\sigma_{xy}$ of a $2+1$ dimensional system of charged particles as one varies some
control parameter.  There are a number of different physical mechanisms and control
parameters that  give rise to this behavior.  In the original, experimentally realized, transition, the system consists of
electrons subject to disorder, with the control
parameter being  an external magnetic field.    A realization that arises naturally in
string theory uses the fact that nonzero transverse conductivities are generated
by integrating out massive fermions.  In field theory terms, the statement is that
one-loop  diagrams for charged fermions induce Chern-Simons terms for external
gauge fields.   Furthermore, the sign of the Chern-Simons terms, and hence the
sign of contribution to $\sigma_{xy}$, is correlated with the sign of fermion mass (recall
that in $2+1$ dimenions a fermion mass term is {\it parity odd}.)  Thus a plateau
transition arises if one (or more) fermion masses is smoothly tuned through zero \cite{Ludwig}.

Such a situation is achieved in string theory by considering various intersecting
brane configurations \cite{Davis:2008nv}.  If the fermions arise from open strings connecting distinct
branes, then the mass is tuned by adjusting the relative separation of the branes
in some mutually transverse direction.  The external gauge field can be taken to be
the gauge fields living on the branes (using a bulk gauge field is also an option in
some cases).

Particular interest attaches to the critical theory right at the transition point where
some fermions are becoming massless.  Geometrically this arises when the intersecting
branes coincide, and there is a nontrivial critical theory living on the intersection.
To apply gauge/gravity duality in this and other contexts, it is convenient
to take the branes carrying the external gauge field (which we'll refer to as the
``electromagnetic field") to be probes living in the near horizon geometry produced
by the other branes.   This is valid provided that the number of probe branes is small,
and indeed we will typically take only a single probe.  In the cases of interest the probes
live in a geometry of the form AdS$_{d+1} \times X$, where at finite temperature
the AdS factor is replaced by a black brane solution.

We then look for a family of probe embeddings labeled by a parameter corresponding
to the brane separation, or equivalently the fermion mass.  Using standard
AdS/CFT methods we can compute the conductivities with respect to the electromagnetic
field.  Here, our main focus is on the critical point itself, achieved when the
probe brane extend all the way down AdS into the throat or horizon, and we will compute
the AC longitudinal and transverse conductivities of the system.

We now describe two specific realizations of these setups.  The impatient reader
could skip to section 3, where we summarize the main results and go on to compute
electrical transport.

\subsection{D3-D7 system}

\label{D3D7section}

The brane construction \cite{Davis:2008nv,rey} contains
$N_7$  D7-branes intersecting $N_3$ D3-branes  over $2+1$ dimensions, summarized by:
\be\label{da}\begin{array}{ccccccccccc} &0&1&2&3&4&5&6&7&8&9 \cr
D3: &\times&\times&\times&\times&      &      &      &      &      &  \cr
D7: &\times&\times&\times&      &\times&\times&\times&\times&\times&
\end{array}
\ee
This intersection has six mixed Neumann-Dirichlet directions, hence it is non-supersymmetric but tachyon free.    The massless sector (when the branes
coincide in $x^9$) consists of $N=N_3 N_7$ complex two-component fermions.
The model focuses on the 2+1 dimensional low energy theory of these fermions.
They have various interaction terms suppressed by powers of the string scale.
With the D3-branes located at $x^9=0$ and the D7-branes at $x^9=L$, the plateau
transition arises by smoothly varying $L$ from  a negative to positive value, or
vice versa.

At strong coupling the system is described by the gravitational dual, which is tractable in the probe approximation \cite{Karch:2002sh}
$N_7 \ll N_3$.   In this limit,  for  large $N_3$ and $gN_3$, one can treat the
D7-branes as probes in the supergravity geometry produced by the D3-branes\footnote{See e.g.
\cite{Erdmenger:2007cm} for a general review on duality constructions with interescting branes.}. The near
horizon geometry at finite temperature is
\be\label{dba}
 ds^2 = {r^2 \over R^2}(-h dt^2+ d\vec{x}^2)+ {R^2 \over r^2}h^{-1}dr^2+R^2 d\Omega_5^2
\ee
with
\be\label{dc}
 h = 1-\frac{r_+^4 }{r^4}~.
\ee
If we write
\be\label{dd} d\Omega_5^2 = d\psi^2 +\sin^2 \psi d\Omega_4^2~,
\ee
then we can identify the  coordinate $x^9$ in (\ref{da}) as
\be\label{de}{ x^9 = r\cos \psi~.}
\ee
The parameters $R$ and $r_+$ are related to the number of D3-branes and the Hawking temperature by the standard formulas
\be\label{df} N_3 = \frac{R^4 }{ 4\pi g \alpha'^2}~,\quad T = \frac{r_+ }{ \pi R^2}~.
\ee

\begin{figure}
\centering
\includegraphics[scale=1]{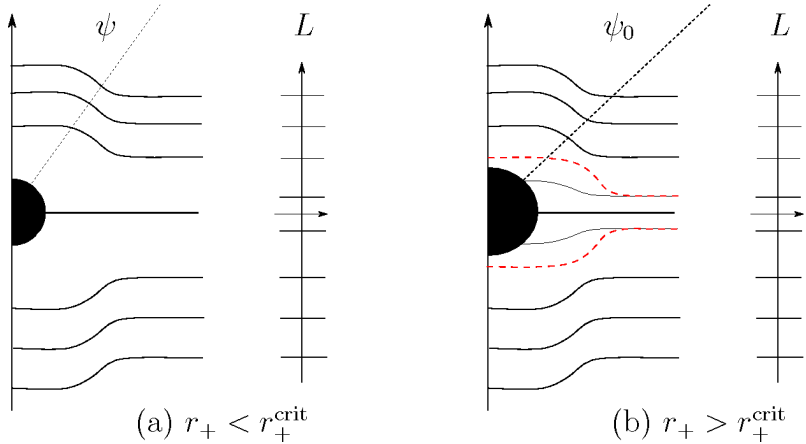}
\caption{Finite temperature probe embeddings (from \cite{Davis:2008nv}).
The vertical axis is $x^9=r\cos \psi$, and the horizontal is $r\sin \psi$.
Each embedding  asymptotes to some particular $x^9$ value,  $x^9=L$.  The black hole embeddings shown in (b) can also be characterized by the angle $\psi_0$ at which they enter the horizon.  See \cite{Davis:2008nv} for discussion of the meaning of $r^{\rm crit}_+$.}
\label{fig3}
\end{figure}

The profile of the probe D7-brane is found by solving the equations of motion derived
from the Born-Infeld action.  The probe lies at some fixed $x^3$ location, say $x^3=0$.
At the boundary we impose the boundary condition $x^9 =L$, but then allow $x^9$ to vary
as we move into the bulk; this captures the interactions between the D3 and D7 branes.
In the angular coordinates we thus allow for a nontrivial profile $\psi(r)$.   At fixed $\psi$ the probe  wraps an $S^4 \subset S^5$, and so our full probe geometry wraps an
$r$ dependent $S^4$.  This $S^4$ either shrinks to zero size at some $r>r_+$, in which case we have a ``Minkowski embedding", or it reaches at finite nonzero size at $r=r_+$, yielding a ``black hole embedding". The different embeddings are sketched in Figure \ref{fig3}.

%kuva alkaa

The probe action consists of two parts: the Born-Infeld action and the Wess-Zumino term which includes the coupling to the background fluxes. The latter contribution will reduce to a Chern-Simons type action, and will be discussed in \ref{CSterm}. We will first focus on the Born-Infeld action. Starting from the $D=7+1$ dimensional Born-Infeld action and assuming that the
worldvolume gauge field has no components tangent to the $S^4$, we can integrate over the
$S^4$ to arrive at a $D=3+1$ dimensional action of the form\footnote{we set $2\pi \alpha'=1$.}
\be\label{dfa} S= -\int\! d^4 x ~\tau(r) \sqrt{-\det(g_4 +F)}~.
\ee
The radius dependent brane tension $\tau(r)$ takes into account the radius dependent
size of the wrapped $S^4$, and is given by
\be\label{dfb} \tau(r) = {N_3 N_7 \over 3\pi^2} \sin^4 \psi(r)~.
\ee

For the purposes of computing electrical conductivities, it is most convenient to
write the induced metric $g_4$ in terms of Eddington-Finkelstein coordinates, as in \cite{Bhattacharyya:2008jc}. To do so,
we need to take into account that the induced metric depends on the profile $\psi(r)$.
We can absorb this dependence by defining the new radial coordinate
\be\label{eca}{ \rt = \int^r_{r_+} \! dx \sqrt{1+h(x)x^2\psi'(x)^2} }~.
\ee
The advanced coordinate is then $v=t+r_*$ with ${d\rt \over dr_*} = {r^2 \over R^2 }h$.
The induced  metric   takes the form
\be\label{ybbb}{ ds^2= 2dvd\rt -U(r)dv^2 +{r^2\over R^2}dx^i dx^i }
\ee
with
\be\label{ybc} U = {r^2\over R^2} h = {r^2\over R^2} -{r_+^4 \over R^2 r^2}~,
\ee
and $r$ understood to be a function of $\rt$ via (\ref{eca}).  The profile function
$\psi(r)$, together with the field strength $F$, can now be computed from (\ref{dfa}).

The preceding discussion  glossed over one important point. As discussed at length in \cite{Davis:2008nv}, the $L=0$ embedding, $\psi = \pi/2$, is unstable, a reflection of the
repulsive D3-D7 force.  We are going to ignore this issue in the remainder of this paper.
One justification is that the instability can be cured by constructions
which lead to essentially the same structure; for instance,  the next example we discuss, probe D6-branes in the ABJM theory, does not suffer from any instability, but has an effective action of the same basic form as (\ref{dfa}).  We expect there to be many
other such examples.\footnote{The paper \cite{Myers:2008me} introduces an additional
flux on the D7-branes; this stabilizes the probe but precludes the possibility
of a QHE transition, since it turns out that the probe must always enter the horizon.}

\subsection{D6-branes in ABJM}

The ABJM gauge theory \cite{Aharony:2008ug} is based on a $U(N)_k\times U(N)_{-k}$ Chern-Simons theory with bifundamental matter, and is dual to IIA string theory on AdS$_4 \times CP^3$.
Massless hypermultiplets transforming in the fundamental  representation  are realized by D6-branes wrapping AdS$_4 \times RP^3$ \cite{Hohenegger:2009as,Gaiotto:2009tk,Hikida:2009tp}.  To obtain a Chern-Simons term for the D6-brane gauge fields we need to add $M$ units of NS-NS 2-form flux, which shifts one of the gauge
group factors to $U(N-M)$, and we also need to give mass $m$ to the hypermultiplets.  This results in a Chern-Simons term proportional to $M$, with a sign correlated with $m$ \cite{Hikida:2009tp}.  Then as above, a plateau transition is realized by a family of massive embeddings that interpolate between opposite sign mass terms.

To be more explicit, the IIA metric is
\be\label{dfd}{ds^2=ds_{AdS_4}^2+4L^2 ds_{CP^3}^2 }
\ee
where $ds_{AdS_4}^2$ refers to a   radius $L$ AdS spacetime (or at finite temperature, a black brane).  Writing the  metric of $CP^3$ as
\bea\label{dfe} ds_{CP^3}^2 &=& d\xi^2 + \cos^2 \xi \sin^2 \xi (d\psi+{\cos\theta_1 \over 2}d\phi_1 - {\cos \theta_2 \over 2}d\phi_2)^2 \\ & +&{1\over 4} \cos^2 \xi (d\theta_1^2+\sin^2 \theta_1 d\phi_1^2)+{1\over 4}\sin^2 \xi (d\theta_2^2+\sin^2\theta_2 d\phi_2^2)
\eea
the massless embedding consists of setting $\xi=\pi/4$, $\theta_1 = \theta_2$, and $\phi_1=-\phi_2$.

Massive embeddings are obtained by allowing for a nontrivial profile $\xi(r)$, just as
in the D3-D7 system above.  Taking into account the background fluxes and integrating
over the compact space, the resulting Born-Infeld actions is of the same form as in (\ref{dfa}) but with
\be\label{dfg} \tau(r) = {2\pi k L^6 } \sqrt{1+b^2 \cos^2 2\xi} \sin 2\xi~.
\ee
Here $b= M/(2kL^2)$.

The induced metric takes the form (\ref{ybbb}) but with
\be\label{dff} U = {r^2 \over L^2}h = {r^2\over L^2}-{r_+^3 \over L^2 r^3}~.
\ee
and the rescaled radial coordinate is now
\be\label{ecab}{ \rt = \int^r_{r_+} \! dx \sqrt{1+4(1+b^2)h(x)x^2\xi'(x)^2} }~.
\ee

To work out the explicit profile $\xi(r)$, and to verify the stability of the massless
embedding, we should use the full action, which contains
the Wess-Zumino coupling to the RR fluxes in addition to the Born-Infeld term.  However, for our computations of AC conductivities we only need the Born-Infeld part.

\subsection{Chern-Simons terms}

\label{CSterm}

A quantum Hall plateau transition corresponds to a change in the DC Hall conductivity  $\sigma_{xy}$.  From the effective action point of view, this corresponds to a change
in the coefficient of the Chern-Simons term for the electromagnetic gauge field. At weak coupling, when
 the CFT description is valid, the Chern-Simons term arises from integrating out the massive fermions. At
 strong coupling, in the
gravitational description, Chern-Simons terms arise from an interaction of the probe
brane with background fluxes.  The corresponding Chern-Simons coefficient depends on the
probe embedding, and can change from one quantized value to another as we evolve through
a family of embeddings.  This is the mechanism used in \cite{Davis:2008nv} to describe
a plateau transition.

After integrating over the compact space and the radial coordinate, the induced Chern-Simons term in our examples takes the form
\be\label{dffa} S_{CS} = {k(\psi_0) \over 4\pi}\int\! A\wedge F~,
\ee
where $\psi_0$ denotes the angle at which the probe brane enters the horizon (this would
be $\xi_0$ for the ABJM example).    As we evolve through a family of embeddings,
$\psi_0$ evolves and induces a transition in  $k(\psi_0)$.     The Chern-Simons
term has the sole effect of  contributing to the DC Hall conductivity as
$\sigma_{xy} = {k\over 2\pi}$.  In what follows, we will be computing contributions to the
AC conductivity from the Born-Infeld part of the probe action, but we should always
remember to add on the Chern-Simons contribution.   When we sit right at the
critical point of the transition there will be no such contribution since the probe
brane embedding is then parity invariant (there could however be a nonzero contribution
from other background probe branes with different emebeddings).

\section{Conductivity for general backgrounds}

Based on the two examples discussed in the previous section, we now introduce a general
class of models to holographically describe quantum Hall plateau transitions and their
critical points.   We take the metric
\be\label{va}{ds^2 =2 dv dr -U(r)dv^2+V(r) dx^i dx^i}
\ee
with $i=1,2$.     At this stage $U(r)$ and $V(r)$ are general functions compatible
with  describing an asymptotically AdS$_4$ black brane metric.\footnote{As before,
the metric takes the form $ds^2 =2 dv d\rt -U(r)dv^2+V(r) dx^i dx^i$,
with $r$  understood to be a function of $\rt$ as in (\ref{ecab}).  But since in this section
we consider arbitrary functions $U$ and $V$, we can can just as well simplify notation and relabel $\rt$ as $r$ and write the metric as in (\ref{va}).   Asymptotically $\rt=r$, and so the
leading large $r$ behavior shown in (\ref{vaa}) is unchanged.}
We thus
impose
\be\label{vaa} U(r) , V(r)  \sim  {r^2 \over R^2}~\quad{\rm as}\quad  r\rightarrow \infty~.
\ee
We assume that there is a horizon at $r=r_+$ at which $U$ vanishes linearly in
$r-r_+$, and $V(r_+)\neq 0$.

As in the examples, we take the  probe action to be a sum of two terms.  First, there is the   Born-Infeld part with a radius dependent tension\footnote{Recall that we are setting $2\pi \alpha'=1$.}
\be\label{vab}{S_{BI} = -  \int\!d^4x~ \tau(r) \sqrt{-\det (g+F)}}
\ee
This radius dependent tension can be thought of as being due to the brane wrapping an $r$ dependent internal space.   At $r=\infty$ the tension goes to a constant; we impose this by writing
\be\label{vac}\tau(r)= f(r)\tau_\infty~,\quad\quad \lim_{r=\infty} f(r)=1~.
\ee
There is also the Chern-Simons term
\be\label{vaba}S_{CS}= {1\over 4\pi} \int\! dk(r)\wedge A\wedge F~.
\ee
The function $f(r)$ arises from the pullback of the bulk fluxes onto the worldvolume.
The Chern-Simons level is given by integrating over the radial direction,
\be\label{vabb}
k= \int_{r_+}^\infty\! dr \p_r k(r) ~.
\ee

As described in the previous section, quantum Hall transitions are realized by
adjusting the transverse separation $L$ between branes. In the current framework this
means adjusting the embedding of the probe, which in turn leads to a one-parameter
family of functions $\tau$, $U$, and $V$.    The parameter $L$ thus controls the
transition, in analogy to how an external magnetic field $B$ controls the more familiar
plateau transition governed by electronic Landau levels.

Even though we do not need to turn on a magnetic field to describe a transition, it is
nonetheless natural to do so, along with a nonzero charge density.  As we will see,
in the absence of these parameters the conductivities turn out to be independent of
frequency, even at finite temperature.    With nonzero charge and magnetic field we find
a nontrivial frequency dependence, allowing us to compare with the model of \cite{sachdev}.
However, it should be kept in mind that these parameters introduce additional length scales
into the problem, while in the model of \cite{sachdev} the only scale is set by the temperature.

The gauge field equations of motion following from (\ref{vab}) admit the following solution:
\be\label{vb}{ F^{(0)}_{12}=B~,\quad\quad F^{(0)}_{rv}=\frac{ \rho(r)/\tau(r) }{ \sqrt{\rho(r)^2/\tau(r)^2+B^2+V(r)^2}}~.}
\ee
We immediately see that $B$ corresponds to an external magnetic field in the boundary
theory.   We have also introduced a radial dependent charge density $\rho(r)$, defined by
\be\label{vba} \rho(r) = {\partial {\cal{L}} \over \partial F_{rv}}~.
\ee
The gauge field equations of motion determine the radial dependence as
\be\label{vbaa} \p_r \rho(r) = {B\over 2\pi} \p_r k(r)~.
\ee
The standard holographic dictionary identifies the boundary charge density as
\be\label{vbab}\rho=\rho(r)|_{r=\infty}~.
\ee

To study the electrical conductivity in linear response we now wish to add a small
electric field and compute the resulting current.  We consider electric fields and currents that are independent of $x^i$, but with a general $v$ dependence.  We break up the gauge field into a
background part  plus a small fluctuation:
\be\label{ZU} A_\mu = A^{(0)}_\mu+   a_\mu~,
\ee
and we'll denote the field strength of $a_\mu$ as $f_{\mu\nu}$.
The boundary electric field is then
\be\label{vda} E_i(v) = \lim_{r\rightarrow \infty} f_{vi}~.
\ee
Note that $v$ plays the role of time on the boundary.

As with the charge density, we introduce a radius dependent current as
\be\label{vdaa} j^i(v,r)=  {\partial {\cal{L}}  \over \partial f_{ri} }~,
\ee
whose radial dependence is given by the $a_i$ equation of motion,
\be\label{vdab} \p_r j^i(v,r) = {1\over 2\pi} \p_r k(r) \epsilon_{ij}E_j~.
\ee
The boundary current is then computed as
\be\label{vdb} j^i(v) = j^i(v,r)|_{r=\infty}~.
\ee
Given our assumptions about the asymptotic form of the metric, this works out to be
\be\label{vca}{  j^i(v)  = \lim_{r\rightarrow \infty}(\frac{r^2}{ R^2} f_{ri}+f_{vi})\tau_\infty~.}
\ee

The entries of the conductivity tensor are defined by
\be\label{vdc} j^i = \sigma_{ij}E^j~.
\ee
Due to rotational invariance we have $\sigma_{xx}=\sigma_{yy}$ and $\sigma_{xy}=-\sigma_{yx}$.
To compute $\sigma_{ij}$ we need to solve the gauge field equations of motion,
subject to the boundary condition (\ref{vda}), and then compute the current from
(\ref{vca}).   In doing so, it is also crucial to impose smoothness of the solution
at the horizon $r=r_+$.   Eddington-Finkelstein coordinates  cover the future
horizon but not the past horizon, so our smoothness condition is really a statement
about the behavior of the solution at the future horizon only.  This condition is
equivalent, but more general, than the condition \cite{Son:2002sd} of imposing ``ingoing boundary conditions" at the horizon.  It is more general since it extends beyond linear response \cite{Bhattacharyya:2008jc}, though this advantage will not be relevant here.

Since we are describing linear response, we only need the action for quadratic
fluctuations of the gauge fields.  The Born-Infeld part is
\be\label{vcb}{S= \frac{1}{ 2 } \int\!d^4x \frac{\tau(r) }{ \sqrt{\Delta_0}}\bigg[2Vf_{ri}f_{vi}-2BF^{(0)}_{rv}\epsilon_{ij} f_{vi}f_{rj}+UVf_{ri}f_{ri}\bigg]}
\ee
with
\be\label{ve}{\Delta_0 = (B^2 +V^2)(1-(F_{rv}^0)^2)=\frac{(B^2+V^2)^2 }{ \rho(r)^2 /\tau(r)^2 +B^2+V^2}~.}
\ee
The Chern-Simons part is
\be\label{veaa}S_{CS}= -{1\over 4\pi} \int\!d^4x \p_r k(r) \epsilon_{ij} a_i f_{vj} ~.
\ee

The nontrivial Maxwell equations take the form
\be\label{vea} \p_r \Big[ \Big(\Ac(r) \delta_{ij}+\Bc(r)\epsilon_{ij}\Big)f_{vi} + \Cc(r)f_{ri} \Big] + \p_v\Big[\Big(\Ac(r)\delta_{ij}-\Bc(r) \epsilon_{ij}\Big)f_{rj} \Big] =-{ k'(r) \over 2\pi \tau_{\infty}}\epsilon_{ij}f_{vj}
\ee
with
\be\label{veb}{\Ac=\frac{f(r) V }{ \ \sqrt{\Delta_0}}~,\quad\quad  \Bc= \frac{f(r)F^{(0)}_{rv}B }{ \sqrt{\Delta_0}}~,\quad\quad \Cc =\frac{f(r)UV }{ \sqrt{\Delta_0}}~.}
\ee
Under our assumptions the large $r$ behavior of these functions is
\be\label{vg}{\Ac \sim 1~,\quad \Bc \sim 0~,\quad \Cc \sim \frac{r^2}{ R^2}~. }
\ee
Choosing the  gauge $a_r=0$ the equations of motion become
\be\label{vga}{ \Cc \p_r^2 a_i +( 2\Ac \p_v + \Cc')\p_r a_i + (\Ac' \delta_{ij}+\Bc'\epsilon_{ij})\p_v a_j=-{ k'(r) \over 2\pi \tau_{\infty}}\epsilon_{ij}\p_v a_j~.}
\ee

Next, we give the components $(a_v,a_i)$ a harmonic time dependence $e^{-i\omega v}$, and
form the complex combinations
\be\label{vgb}a_\pm = a_x \pm i a_y~,
\ee
leading to
\be\label{vf}{\Cc \p_r^2 a_\pm +(\Cc' -2i \omega\Ac )\p_r a_\pm -i\omega (\Ac' \mp i\Bc')a_\pm =\pm { k'(r) \over 2\pi \tau_{\infty}}\omega a_\pm~.}
\ee

Asymptotically AdS solutions will have asymptotic behavior
\be\label{vgx}{a_\pm = a^{(0)}_\pm +\frac{R^2}{ r} a_\pm^{(1)}+\cdots }
\ee
This gives an electric field and current
\be\label{vh}{ E_\pm = f_{v\pm}= -i\omega a_\pm^{(0)}~,\quad\quad j_\pm =\tau_\infty(-a^{(1)}_\pm -i\omega a^{(0)}_\pm )~.}
\ee
The conductivities are therefore
\be\label{vi}{\sigma_{\pm}(\omega) =\bigg(1-i\frac{a^{(1)}_\pm }{ \omega a_\pm^{(0)}} \bigg)\tau_\infty~.}
\ee
In $xy$ coordinates we have
\be\label{via} \sigma_{xx}= \sigma_{yy} ={1\over 2}(\sigma_+ + \sigma_-)~,\quad\quad
\sigma_{xy}= -\sigma_{yx} ={i\over 2}(\sigma_+ - \sigma_-)~.
\ee

What we've computed above  is the contribution to the conductivity from the Born-Infeld part of the probe
action.  As discussed in section (\ref{CSterm}), in general we should also add in the
contribution from the Chern-Simons term on the probe, which shifts the DC Hall conductivity by the amount $\sigma_{xy}={k\over 2\pi}$.

Computation of the AC conductivity has now been reduced to solving (\ref{vf}).  In general, this requires numerical analysis and specific choices for the functions
$\tau$, $U$, and $V$.   In particular limits we can solve the problem analytically, as we discuss in the next few sections.   In the following we will always be assuming that
the probe is described by a black hole embedding; this is necessary in order for
the dissipative part of the conductivity to be nonzero, since for a Minkowski embedding
there is nowhere for energy to flow out of the system (at least in the probe approximation).

\subsection{DC limit}

First consider the $\omega\rightarrow 0$ DC limit. In this limit we can compute
the conductivities in the general case.    This includes the DC conductivity at and away
from the critical point for our entire class of theories. A similar computation was given in
\cite{O'Bannon:2007in}.
Looking at (\ref{vi}), we see that
to compute the DC conductivity we need to compute $a_\pm$ to first order in $\omega$.

At zeroth order the equation of motion (\ref{vf}) becomes
\be\label{vj}{\p_r  (\Cc \p_r a_\pm)=0~. }
\ee
Assuming there is a horizon where $U$ vanishes linearly in $r-r_+$, the only smooth solution
is $a_\pm = c_\pm$, a constant.   Plugging this in, we have at the next order the equation
\be\label{vk}{ \p_r  (\Cc \p_r a_\pm)=i \omega \p_r (\Ac \mp i\Bc\mp i {k(r)\over 2\pi \tau_\infty})c_\pm }
\ee
the smooth solution to which is
\be\label{vl}{ \p_r a_\pm(r) = \frac{i\omega }{ \Cc(r)}\bigg[ \Big(\Ac(r)\mp i \Bc(r)\pm i{k(r)\over 2\pi \tau_\infty}\Big)- \Big(\Ac(r_+)\mp i\Bc(r_+)\mp i{k(r_+)\over 2\pi \tau_\infty}\Big)\bigg]c_\pm~. }
\ee
This has large $r$ behavior
\be\label{vm}{ \p_r a_\pm(r) \sim \frac{i\omega R^2 }{ r^2 }\bigg[ 1- \big(\Ac(r_+)\mp i\Bc(r_+)\big)\mp i{k\over 2\pi \tau_\infty}\bigg]c_\pm}
\ee
and so we read off
\be\label{vn}{a^{(1)}_\pm = -{i\omega   }\bigg[ 1- \big(\Ac(r_+)\mp i\Bc(r_+)\big)\mp i{k\over 2\pi \tau_\infty}\bigg]c_\pm~.}
\ee
Using that  $a^{(0)}_\pm = c_\pm+ {\cal O}(\omega)$,  we find the DC conductivity
\be\label{vo}{\sigma_\pm = \bigg(\Ac(r_+) \mp i\Bc(r_+) \bigg)\tau_\infty\mp i {k\over 2\pi}~,}
\ee
or,
\bea\label{vp} &&\sigma_{xx}=\sigma_{yy}=  \Ac(r_+)\tau_\infty = \frac{\sqrt{\rho(r_+)^2/\tau(r_+)^2+B^2+V(r_+)^2} }{ B^2+V(r_+)^2}V(r_+)\tau(r_+)  \nonumber \\
 &&\sigma_{xy}=-\sigma_{yx}= \Bc(r_+)\tau_\infty = \frac{\rho(r_+) B }{ B^2+V(r_+)^2~}+{k(\psi_0)\over 2\pi}~.
\eea
In the last line we have indicated that the Chern-Simons level $k$ depends on the angle
$\psi_0$ at which the probe enters the horizon.  From (\ref{vbaa}) the quantity
$\rho(r_+)$ is related to the actual charge density as
\be\label{vpz} \rho(r_+)= \rho -{kB \over 2\pi}~.
\ee
Taking $k=0$ we can check that our result agrees with that in \cite{O'Bannon:2007in}.

The result (\ref{vp}) has some interesting features.  First, the conductivities are
independent of the metric function $U$.  In fact, the only dependence on the original
black brane metric is through $V(r_+)$, which is a measure of the entropy density of the
black brane.  In particular, for a probe embedded in an asymptotically $d+1$ dimensional
black brane geometry, the entropy density is proportional to $\big(V(r_+)\big)^{d-1\over 2}$.   The only other theory dependent parameter is the brane tension at the horizon,
$\tau(r_+)$.  So we see that the DC conductivities  are nearly universal functions of
$\rho$, $B$ and the entropy density, with only one additional parameter, $\tau(r_+)$.
This is an interesting holographic prediction for a large class of critical points.

One interesting special case is $\tau(r_+)=0$, which gives
\bea\label{vq} &&\sigma_{xx}=\sigma_{yy}=\frac{\rho V(r_+) }{ B^2+V(r_+)^2}  \nonumber \\
 &&\sigma_{xy}=-\sigma_{yx} = \frac{\rho B }{ B^2+V(r_+)^2}
\eea
This value of $\tau(r_+)$ corresponds to a critical embedding  which just touches the
horizon.  In can be thought of as the special case for which the black hole and Minkowski family of embeddings meet.   It represents a  metal-insulator transition, as can be seen from the fact that the conductivity vanishes in the absence of a net charge density.  What is interesting is that the  conductivities here are universal
functions of $\rho$, $B$, and the entropy density.

\subsection{Small $(B,\rho)$ expansion at the critical point}

At $B=\rho=0$ the AC conductivity turns out to be independent of $\omega$ as a consequence of electric-magnetic duality in the bulk \cite{Hartnoll:2007ip}.   For general
$(B,\rho)$ we require numerical analysis to determine the $\omega$ dependence.  It is useful to consider an expansion to first nontrivial order in $(B,\rho)$ in order to obtain an analytical expression with a nontrivial $\omega$ dependence.  To make analytical progress we will also assume that we are at the critical point of the transition, which
implies that $f(r)=1$, and hence constant $\tau$; recall that this means that the probe is then wrapping a fixed internal space, not depending on $r$.  Also, recall that the
Chern-Simons term is absent at the critical point.

Setting $B=\rho=0$ in (\ref{vf}) the solution   for $a_\pm$ is simply a constant.
Since the conductivity (\ref{vi}) is independent of the overall normalization of $a_\pm$, we can choose $a_\pm=1$ at this order, and also impose this value as a boundary
condition at the horizon at every order in $\omega$.

At the next order we write
\be\label{vqa}a_\pm = 1+b_\pm~,
\ee
with the boundary condition $b_\pm(r_+)=0$ as explained above.
We expand in the parameter $\epsilon \sim(B,\rho)$, and we readily find  $b_\pm \sim \epsilon^2$.
The expansions for the coefficient functions are
\begin{equation}
\begin{split}
  \Ac &= f(r)\left[1+\frac{1}{2V(r)^2}\left(\rhot^2-B^2\right)\right]+\cdots  \\ \quad\quad \Ac'\mp i \Bc' &= f'(r)-\frac{f'(r)}{2V(r)^2}(\rhot^2+B^2)+\frac{f(r) V'(r)}{V(r)^3}(B\pm i \rhot)^2+ \cdots~,
%,\\ \quad\quad
%\Cch &= U(r)\Ac
\end{split}
\end{equation}
where we are writing   $\rhot = \rho/\tau$.
Substituting in, we find that at this order $b_\pm$ obeys the equation
\begin{equation}
\left(U b'_\pm -2 i \omega b_\pm \right)'= -{ i \omega\over 2} (B\pm i \rhot)^2 \left( \frac{1}{V^2}\right)'~.
\end{equation}

We integrate twice to get
\begin{equation}
 b_\pm(r) = {i \omega\over 2} (B \pm i \rhot)^2  \int_{r_+}^r \! {dx\over U(x)}  \left({1\over V(r_+)^2}-{1\over V(x)^2}\right)  e^{2 i \omega\int_x^r \! {dy \over U(y)}}~,
\end{equation}
where we imposed  the boundary condition $b_\pm(r_+)=0$~.

{}From this we can read off the asymptotic behavior of the full solution to the required order:
\be
\label{ye} a_\pm(r) = 1 +b_\pm(\infty) -2i \omega \left( b_\pm(\infty)+{(B\pm i \rhot)^2\over 4V(r_+)^2} \right)\frac{R^2}{ r}+ \cdots
\ee
This finally gives the AC conductivity to order $\epsilon^2$:
\begin{equation}
\label{yg} \sigma_\pm(\omega)  = \tau -\frac{\tau}{ 2} (B\pm i \rho/\tau)^2G(\omega)
\end{equation}
with
\be\label{yiz} G(\omega)= {1\over V(r_+)^2} +2 i \omega  \int_{r_+}^\infty {\mathrm{d} x\over U(x)} \left({1\over V(r_+)^2}-{1\over V(x)^2}\right)  e^{2 i \omega \int_x^\infty \! {dy \over U(y)}}~.
\ee
Equivalently:
\bea\label{yi} \sigma_{xx}(\omega) &=& \tau +\frac{1}{ 2}\tau (\rho^2/\tau^2 - B^2)G(\omega)\nonumber \\
 \sigma_{xy}(\omega) &=&  \rho B  G(\omega)~,
\eea

Let us comment on the regime of validity of this result.  In solving the differential
equation we kept terms of order $\omega \epsilon^2$ and discarded terms of order
$\omega^2 \epsilon^4$.  Thus we actually need $\omega \epsilon^2$ to be small, and not just $\epsilon$ itself.   Hence, for fixed small $(B,\rho)$ the result is valid for large,
but not arbitrarily large, frequencies.

\begin{figure}
\centering
\includegraphics[scale=0.5]{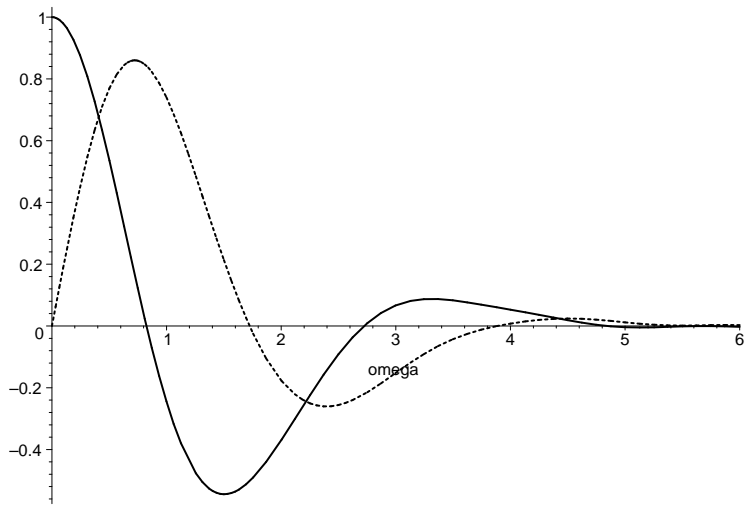}
%{F4.eps}

\caption{The real (solid) and imaginary (dashed) parts of $G(\omega)$. Note the resemblance
of the real part to Fig. \ref{fig1}.}
\label{fig4}
\end{figure}

Above we show a plot (Fig. \ref{fig4}) of the real and imaginary parts of $G(\omega)$ in the case corresponding to the D3-D7 system (and
setting $r_+=R=1$ by a rescaling), for which $U=r^2- {1\over r^2}$ and $V=r^2$. %{\bf{INSERT PLOT.}}

For small
$(B,\rho)$ this leads to excellent agreement with our general numerical results.
We see that $G(\omega)$ exhibits oscillations on a scale set by the temperature,
$\Delta \omega \sim r_+ \sim T$.

\subsection{High frequency limit and universal conductivity}

We now turn our attention to
the high-frequency limit $\omega/T \gg 1$ of the conductivity. We first make some general comments to put our results in context.   A condensed
matter system at a quantum critical point is described at large spacetime length scales by  a scale invariant quantum field theory \cite{Sach}.  Within this regime, finite temperature can be introduced into the quantum field theory, as can additional large length scales
corresponding to magnetic fields and charge densities.  One thus has a finite
temperature scale invariant QFT deformed by various operators and charge densities, which
is the description that AdS/CFT seeks to make contact with.

If we sit  right at the fixed point but turn on finite temperature, the AC conductivity can only be a function of $\omega/T$. As discussed in, {\em e.g.},  \cite{damlesachdev} the functional dependence on $\omega/T$ is  universal in this regime, with any non-universality suppressed by powers of microscopic length scales.  It follows that large $\omega$ is equivalent to small $T$, and that the conductivity in the limit $\omega/T \rightarrow \infty$ defines a universal number in dimensionless units.  This number
is to be distinguished from the DC conductivity at the transition, which is defined
by the limit $\omega/T \rightarrow 0$.

Similarly, we can compute or measure the limiting values of the conductivity in the
presence of nonzero magnetic fields and charge densities.  These describe relevant deformations of the theory, and so are expected to have an important affect at low
frequencies but not at high frequencies.

We can make these points precise.  We already computed the general DC conductivity in
(\ref{vo}).    As seen from (\ref{vp}), it is purely a function of quantities
evaluated at the horizon, which accords with general intuition on the relation between AdS radial position and boundary length scales.
The conductivity in the limit $\omega/T \rightarrow \infty$ is also easily computed.
Examining  (\ref{vf}) we see that $a_\pm$  is a well defined solution in the
limit, and then from (\ref{vi}) we have that
\be\label{ZZ} \lim_{{\omega / T}\rightarrow \infty} \sigma_\pm(\omega/T) = \tau_\infty~. \ee
So in the limit the conductivity is determined purely by the asymptotic brane tension. Note that this result holds at nonzero $(B,\rho)$, confirming the intuition that these
are relevant deformations, and hence unimportant in the UV.

It is also interesting to discuss the subleading behavior as $\omega/T \rightarrow \infty$.   Interestingly, this can be determined explicitly for our general class
of models at the critical point.  We will take $R=1$ by rescaling $r$ to simplify
some expressions.  We proceed by solving the differential equation (\ref{vf}) as
a power series in $1/\omega$.       We write
\be\label{vva} \ah_\pm = b_0 + {b_1 \over \omega}+ {b_2 \over \omega^2} + \cdots
\ee
At lowest order we find
\be\label{vvc} b_0(r) = {c_0\over \sqrt{\Ac(r)}}e^{\pm{i\over 2}\int_{r_+}^r \! d\rt {\Bc'(\rt)\over \Ac(\rt)}}~.
\ee
The integration constant $c_0$ can be chosen arbitrarily since it cancels out of the
conductivity.   We choose it so that $b_0(\infty)=1$.

Substituting in the expansion (\ref{vva}) into (\ref{vf}), we obtain the recursion relation
\be\label{vvb} 2 \Ac b'_{n+1} +(\Ac'\mp i\Bc') b_{n+1}= -i (\Cc b_n')'~,
\ee
which we solve as
\be\label{vvd} b_{n+1}(r) = -{i\over 2\sqrt{\Ac(r)}} \int_{r_+}^r \! d\rt~ {\Cc(\rt) b'_n(\rt) \over \sqrt{\Ac(\rt)}} e^{\mp {i\over 2}\int_{r}^{\rt} \! d\hat{r} {\Bc'(\hat{r})\over \Ac(\hat{r})}}~.
\ee
The free integration constant can again be chosen at will, since it can be absorbed into $c_0$. From the formula (\ref{vi}) for the conductivity, we know that a subleading contribution at large frequency  arises from a nonzero ${1\over r}$ term in the large $r$ expansion of
$a_\pm$.    By differentiating (\ref{vvd}) at large $r$ we find the large $r$ relation
$b'_{n+1} =-{i\over 2}(r^2 b'_n)' + \ldots$.     From (\ref{vvc}) together with
the expressions (\ref{veb}) we find that $b'_0=-(B\pm i \rho/\tau)^2{1 \over r^5}+\ldots$.  Together, this implies that we first get a ${1\over r}$
contribution at order $b_3$, which behaves for large $r$ as
\be\label{vve} b_3 = {\rm const}  -{3i (B\pm i \rho/\tau)^2  \over 4r} + \cdots
\ee
We then read off the large $\omega$ behavior of the conductivity as
\be\label{vvf} \sigma_\pm = \bigg(1 - {3\over 4} {(B\pm i \rho/\tau)^2 \over \omega^4}+\cdots \bigg)\tau~.
\ee
This result is completely universal, in that it is independent of the precise forms
of the original coefficient functions $U$ and $V$.   From familiar UV/IR reasoning it is not surprising that the leading high frequency behavior of the conductivity only uses
the asymptotic AdS behavior of the geometry, but here we see that this is also true
of the subleading behavior.
We have checked that (\ref{vvf}) indeed matches on to our numerical results (presented in the next section)  at large
$\omega$ to excellent accuracy.

The asymptotic result for the conductivity is
temperature independent, since the factors of $r_+$ all canceled out. This suggests that we should be able
to understand the subleading term as a property of correlation functions in the zero
temperature theory.  We can consider this problem from the point of view of the boundary theory (see also
\cite{Myers:2008cj,Myers:2007we} for related discussion).
The conductivity is related to the retarded current-current correlator according
to the Kubo formula:
\be\label{hfk}{\sigma_{ij} (\omega) = \frac{i}{ \omega}C_{ij}(\omega,0) \ ,}
\ee
with
\be\label{hfl} C_{ij}(\omega ,q) = \int \!d^3x e^{-i\omega t + iqx} \langle [J_i (t,x),J_j (0))]\rangle \ ,
\ee
and
where the expectation value is calculated in thermal equilibrium at temperature $T$.
Since we are considering a spatially homogenous electric field we set $q=0$.

Now,  $J_i$ is an operator of scale dimension $2$, and so $C_{ij}(\omega)$ has scale
dimension $1$.  At  $T=B=\rho=0$ scale invariance thus fixes $C_{ij}(\omega)\sim \omega$,
and hence $\sigma_{ij}(\omega) \sim \omega^0$, which is just the usual statement that
conductivity is dimensionless in $D=2+1$.

We now break the scale invariance by turning on nonzero temperature and magnetic field,
which have scale dimensions $\Delta_T =1$, $\Delta_B = 2$.  We are keeping
$\rho=0$ for simplicity.    At some order, the large
$\omega$ expansion of $C_{ij}$ will involve both scales, but  we can take as an
{\it ansatz} that the first subleading term is $T$ independent.  Focussing on
$\sigma_{xx}$, parity forces $B$ to appear in even powers, and hence the allowed
subleading behavior corresponds to
\be\label{ZX} \sigma_{xx} \sim \omega^0  +\sharp{ B^2 \over \omega^4} + \cdots
\ee
Computing the coefficient of the subleading term requires knowledge of a four-point
function evaluated at the fixed point, but    the fact that the form of
(\ref{ZX}) agrees with our previous result (\ref{vvf}) seems to justify the assumption of temperature independence at this order.

\section{AC conductivity of finite temperature critical point}

We now turn to the computation of the conductivity for generic $B$, $\rho$, and
$\omega$.   To keep the numerics tractable we will only consider the critical point,
which corresponds to setting $f(r)=1$, and hence $\tau$ is a constant.  For definiteness,
we also focus on the particular example of the D3-D7 system reviewed in section (\ref{D3D7section}). For this case, the problem to be solved is summarized as follows.  We need to solve the equation (\ref{vf}),
\be\label{vfz}{\Cc \p_r^2 a_\pm +(\Cc' -2i \omega\Ac )\p_r a_\pm -i\omega (\Ac' \mp i\Bc')a_\pm =0~,}
\ee
where the coefficient functions are given in (\ref{veb}) with
\be\label{vfy} f(r)=1~,\quad\quad U= {r^2 \over R^2}-{r_+^4 \over R^2 r^2}~,\quad\quad V={r^2 \over R^2}~.
\ee
We look for a solution with
asymptotic behavior
\be\label{vgz}{a_\pm = a^{(0)}_\pm +\frac{R^2}{ r} a_\pm^{(1)}+\cdots }
\ee
and then read off the conductivity from
\be\label{viz}{\sigma_{\pm}(\omega) =\bigg(1-i\frac{a^{(1)}_\pm }{ \omega a_\pm^{(0)}} \bigg)\tau~.}
\ee

We can do some rescalings to simplify.   We first set $R=1$, as can obviously be done
by a rescaling of $r$.    To scale out $r_+$ we first introduce
\be\label{av} x= r/r_+
\ee
so that the horizon is at $x=1$.  Also define
\be\label{aw} \Bh = \frac{B}{ r_+^2}~,\quad \rhoh =\frac{\rho }{ \tau r_+^2}~,\quad \omh = \frac{\omega }{ r_+}~,\quad \ah_\pm = {r_+ a_\pm }~.
\ee
We can then write
\be\label{ax}{ \Ac = \frac{x^2 \sqrt{x^4 +\Bh^2 +\rhoh^2 }}{ x^4 +\Bh^2}~,\quad \Bc= \frac{\Bh \rhoh }{ x^4 +\Bh^2}~,\quad \Cc = r_+^2 (x^4-1) \frac{\sqrt{x^4 +\Bh^2 +\rhoh^2 }}{ x^4 +\Bh^2} \equiv r_+^2 \Cch }
\ee
and our equation (\ref{vfz} becomes
\be\label{axa}{ \Cch  \ah_\pm'' +(\Cch'-2i\omh \Ac) \ah_\pm' -i\omh(\Ac' \mp i \Bc')  \ah_\pm = 0}
\ee
where now $'=\frac{d}{ dx}$.

The asymptotic behavior is now
\be\label{bh}{ \ah_\pm =  \ah_{\pm}^{(0)} + \frac{1}{ x} \ah_{\pm}^{(1)}  + \cdots }
\ee

and the conductivity is
\be\label{bl} \sigma_{\pm}(\omh) =  \bigg(1 -i \frac{\ah_\pm^{(1)}}{ \omh \ah_\pm^{(0)}} \bigg)\tau \ .
\ee

Given the above, we will compute the conductivity as a  function of the rescaled frequency $\omh$  and the rescaled parameters $\rhoh$ and $\Bh $.

\subsection{Numerical results}

We are now ready to numerically integrate (\ref{axa}).
We  demand that $\ah_\pm$ is a smooth function at the horizon $x=1$, which we recall
is actually a statement about smoothness at the {\it future} horizon, since Eddington-Finkelstein coordinates do not cover the past horizon.
Since only the ratio $\frac{\ah_\pm^{(1)}}{ \omh \ah_\pm^{(0)}}$ appears in the
conductivity, we are free to choose the horizon boundary condition $\ah_\pm(1)=1$.

If we evaluate (\ref{axa}) at the horizon, where $\Cc(1)=0$ and $\Cc'(1)=4\Ac(1)$, we
can solve for the initial value of  $\ah'_\pm(1)$ as
\be\label{acasz}
 \hat{a}'_{\pm}(1) =
\frac{i\hat{\omega}\big(\Ac'(1)\mp i\Bc'(1)\big)}{ ( 4-2i\hat{\omega}) \Ac (1)} \ .
\ee
Our initial conditions at the horizon are thus fixed, and we can proceed to integrate
outward in a straightforward fashion.

\begin{figure}[h!bt]
\begin{center}
\noindent
\includegraphics[width=0.7\textwidth]{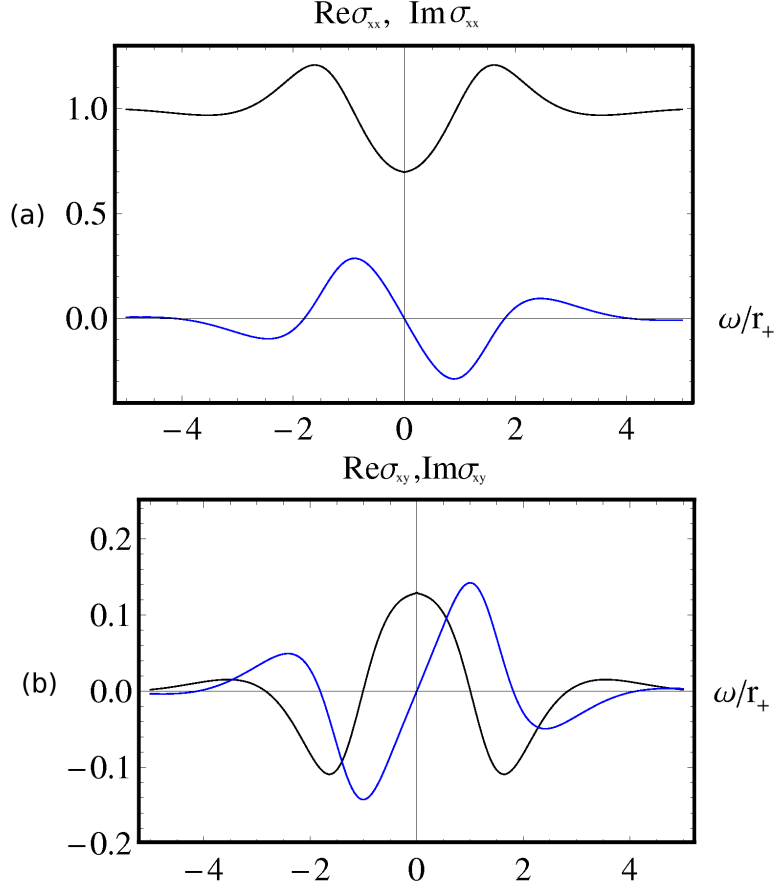}
\end{center}
\caption{Real and imaginary parts of the (a) longitudinal and (b) Hall conductivity, plotted over a range of negative to positive frequencies to illustrate symmetry and antisymmetry.
In (a), the top (black) curve is the real part and bottom (blue) curve is the imaginary part of the longitudinal conductivity
$\sigma_{xx}$, for parameter values $\rhoh=0,\Bh=1,\tau=1$. The value at $\omega=0$ corresponds to the DC conductivity $\sigma_{xx}=1/\sqrt{1+\Bh^2}$, a special case of (\ref{vp}). In (b), the symmetric (black)
curve is the real part and the antisymmetric (blue) curve is the imaginary part of the Hall conductivity
$\sigma_{xy}$, for parameter values $\rhoh=0.25,\Bh=1,\tau=1$.}
\label{fig5}
\end{figure}

Figure \ref{fig5} illustrates the symmetry properties of the conductivities. The real part of the
longitudinal and Hall conductivity is symmetric under the reflection $\omega \rightarrow -\omega$, while
the imaginary parts are antisymmetric.  The figure shows
the longitudinal conductivity $\sigma_{xx}$ at $\rhoh=0,\Bh=1$. A non-zero Hall conductivity requires
$\rhoh\neq0$, the figure shows a plot of $\sigma_{xy}$ for $\rhoh=0.25,\Bh=1$. The third parameter, the Born-Infeld action coefficient $\tau$ corresponds to rescaling of the $y$-axis, and determines the
asymptotic value of ${\rm Re} \sigma_{xx}$. Here we have set $\tau=1$.

%kuva alkaa
\begin{figure}[tbp]
\includegraphics[scale=1]{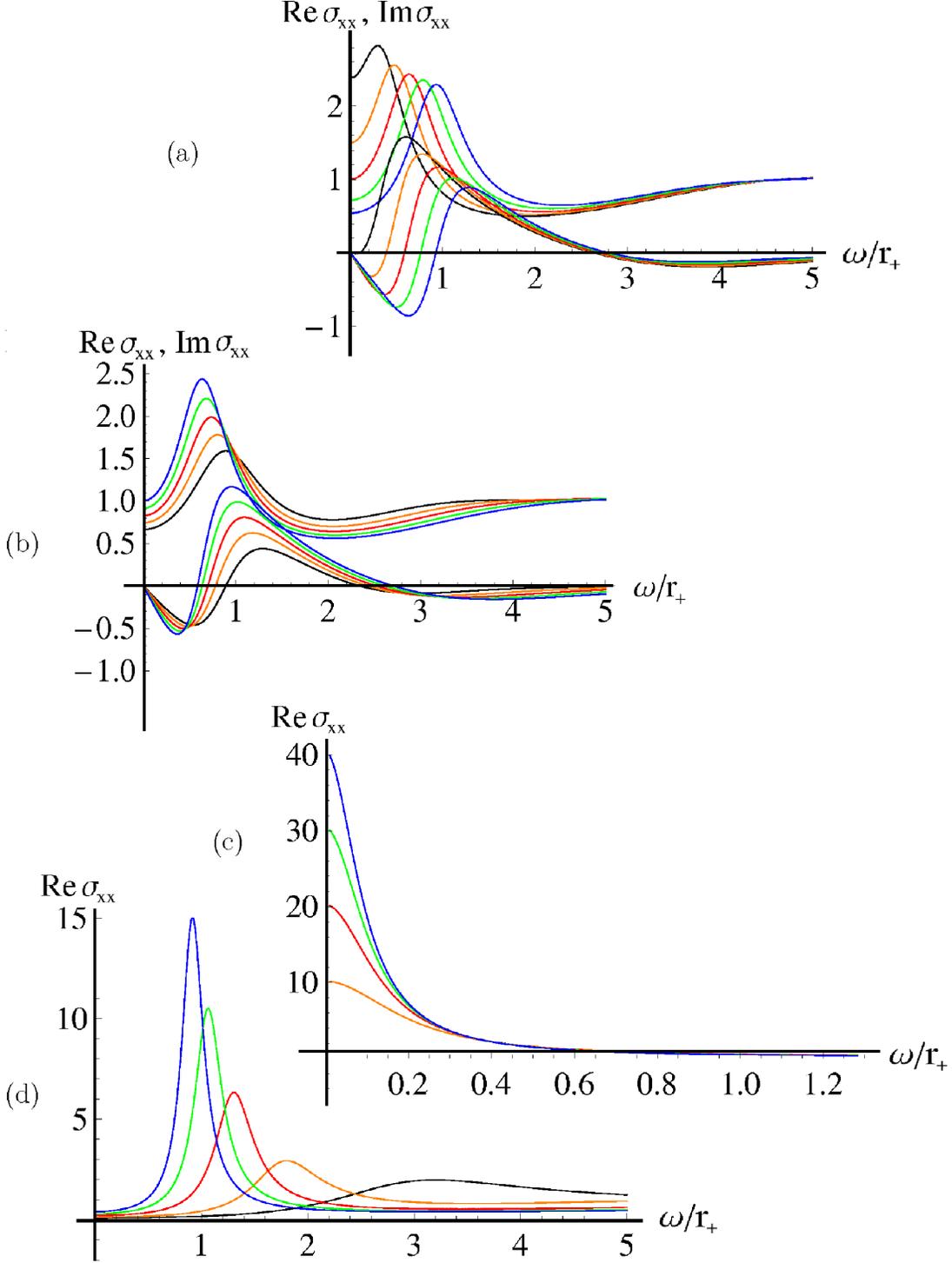}
\caption{Real part (top curves) and imaginary part (bottom curves) of the longitudinal conductivity
$\sigma_{xx}$. The $x$-axis is $\omega/r_+$.
Figure (a)
%is for $\rhoh=1$,  the curves represent 5 different values
%$\Bh=0,0.5,1,1.5,2$, with colors black, orange, red, green, blue.
is for $\rhoh=4.5$,  the curves represent 5 different values
$\Bh=1,1.5,2,2.5,3$, with colors black, orange, red, green, blue.
(b) is for $\Bh=2$, with $\rhoh=2.5,3.0,3.5,4.0,4.5$,
(c) is for $\Bh=0$, with $\rhoh=0,10,20,30,40$,
(d) is for $\Bh=10$, with $\rhoh=0,10,20,30,40$.
%$B=0$ curves
%are straight lines, then the curves develop peaks and valleys, $B=1$ giving the largest amplitude
%curves.
The values at $\omega=0$ correspond to the DC conductivity $\sigma_{xx}$ of (\ref{vp}).}
\label{fig6}
\end{figure}

As the parameter $\rhoh$ is turned on, the plots develop more structure. Figure \ref{fig6} depicts plots
of $\sigma_{xx}$ for various values of the parameters $\rhoh,\Bh$.
%The parameter $\tau$
%(the coefficient
%of the Born-Infeld action)
%has been set to one, as it simply sets the overall scale of the $y$-axis.
Figure \ref{fig6} (a) depicts the effect of keeping $\rhoh$ fixed while $\Bh$ is varied: the peaks
of the curves move towards higher frequencies, as would be expected if they are associated with a
cyclotron frequency.  Figure \ref{fig6} (b) illustrates the opposite choice, keeping $\Bh$ fixed while $\rhoh$ is varied. The (Drude) peak of the conductivity grows when charge density increases, as expected.
Figures \ref{fig6} (c) and (d) focus on the real part of the longitudinal conductivity, showing the effect
of larger charger densities for two fixed values of the magnetic field. When the magnetic field is
turned on, (d) illustrates how the Drude peak grows but also moves to lower frequencies as the charge
density increases. In all figures the real part of $\sigma_{xx}$ asymptotes to one, because
we fixed $\tau=1$.

\begin{figure}
\includegraphics[scale=1]{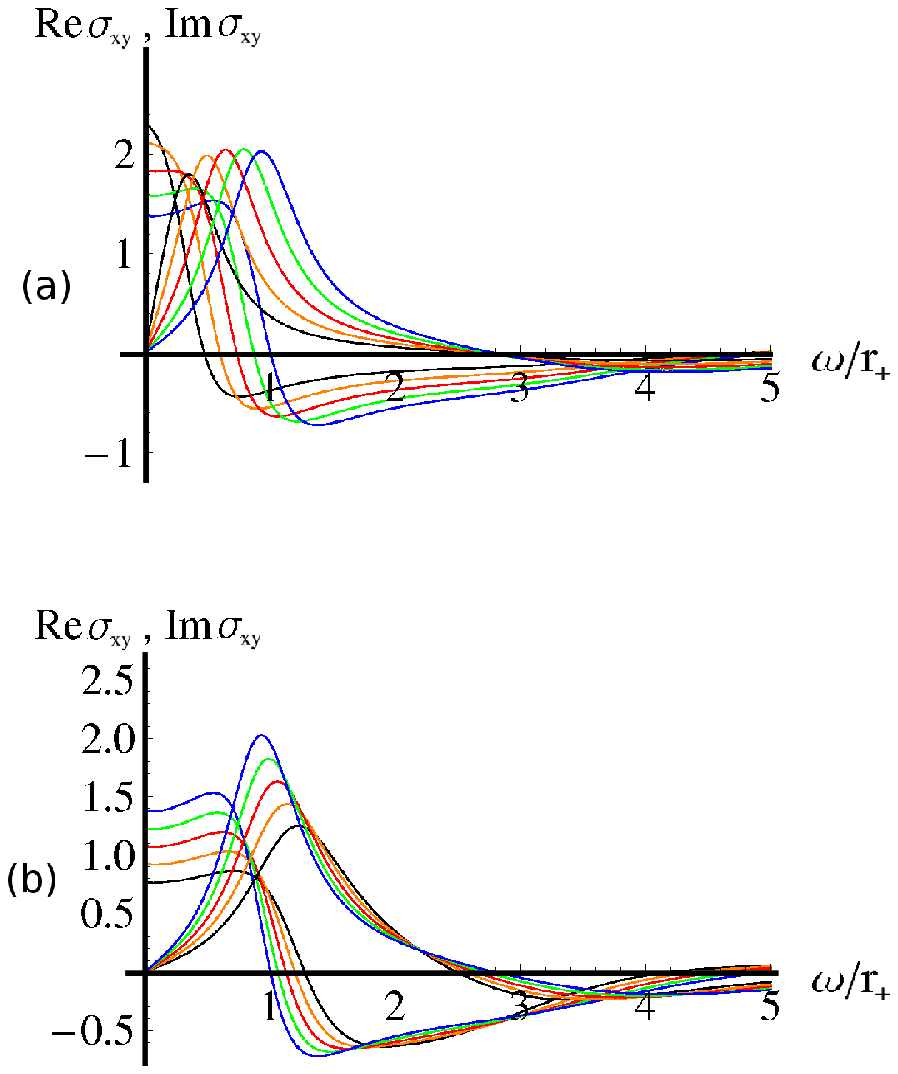}
\caption{Real part (non-zero values at $\omh=0$)
and imaginary part (passing through zero at $\omh=0$) of the Hall conductivity
$\sigma_{xy}$.
Figure (a) is for fixed  $\rhoh=4.5$, with $\Bh=1,1.5,2,2.5,3$, plot (b)
is for fixed $\Bh=3$, with $\rhoh=2,2.5,3,3.5,4,4.5$. }
\label{fig7}
\end{figure}

For the Hall conductivity, the plots are qualitatively similar, except that now the high-frequency limit
gives zero. Figure \ref{fig7} shows two series of plots for the real and imaginary part of $\sigma_{xy}$. %Fig. \ref{fig7} (a) displays their symmetry and antisymmetry.
The figure (a) shows the effect of
varying $\Bh$ while $\rhoh$ is fixed, and (b) depicts the opposite case.

%kuva alkaa

In Fig. \ref{fig2} we give a comparison with the results of Sachdev. Note that many of the plots in Figures 6 
and 7 are qualitatively similar to Figure 2 of \cite{Hartnoll:2007ip}. This suggests that the first peak
may be determined by a cyclotron resonance, which in turn depends on kinematics \cite{Hartnoll:2007ih}.
However, for the agreement depicted in Fig. \ref{fig2},  we need to 
purposely choose three parameters $(\rho,B,\tau)$ in order to make the curves match up nicely.
It is unexpected that we can do this, since we are comparing four different curves while only tuning three constants. The cyclotron resonance may not be sufficient to explain the more detailed agreement.

\section{Conclusion}

Quantum Hall critical points are a very promising arena for applying AdS/CFT methods to
condensed matter physics.  Here we studied the AC electrical response in a class of such critical points,  and obtained a variety of analytical and numerical results.  One
outcome was the unexpectedly good agreement between our results and those
obtained previously by Sachdev, as displayed in Fig. \ref{fig2}. We note that the curves match nicely despite
of the utterly different computational methods. It would of
course be especially interesting to compare with  experimental results for AC transport
in these systems, if/when these become available.

Also encouraging was the high degree of universality exhibited by our results.  This was
seen in both the low and high frequency regimes.  In the DC limit, we found that the
conductivities over the entire plateau transition just depend on the entropy density of
the theory and the effective brane tension at the horizon.  In the high frequency regime
we found universal results for both the leading and subleading behavior at the critical point.
It is of course precisely such robust quantities that one might hope to compare successfully with real physical systems.

We have emphasized an interpretation of our critical points in terms of the quantum Hall effect, but our results can also be viewed in more general terms as modeling
$2+1$ dimensional critical points with nonzero charge density and magnetic field.  In this sense, our probe brane based approach is complementary to that based on using the dyonic
black brane solution as a starting point \cite{Herzog:2007ij,Hartnoll:2007ai,Hartnoll:2007ih,Hartnoll:2007ip}.
It might be informative to generalize further by combining the two sets of ingredients. In the near
 future we plan to study the poles in the complex frequency plane that determine the response curves,
 especially the role of the cyclotron resonance.

\bigskip

{\large \bf Acknowledgments}  We thank  Josh Davis, Ari Harju, Sean Nowling, Gordon Semenoff, and Patta Yogendran for
useful comments and discussions. JA has been supported in part by the Magnus Ehrnrooth foundation. EK-V has been supported in part by the the Academy of
Finland grant nr. 1127482.  Work of PK is supported in part by NSF grant PHY-0456200. VS-U has been supported in part by the V\"{a}is\"{a}l\"{a}\ fund of the Finnish Academy of Science and Letters.
EK-V thanks the Aspen Center for Physics for hospitality during the completion of this work.

\end{document}